\documentclass[aps,pra,floatfix,superscriptaddress,twocolumn,showpacs]{revtex4}
\usepackage{amsfonts}
\usepackage{amssymb}
\usepackage{amsmath}
\usepackage{graphicx,psfrag}
\usepackage{bm}
\usepackage[T1]{fontenc}
\usepackage{calligra}
\usepackage{dcolumn}    
\usepackage{epsf}  
\usepackage{epsfig}      
\usepackage{times}  
\usepackage{nicefrac}  
\usepackage{color}      
  
\newcolumntype{.}{D{x}{}{-1}}  
  
\newcommand{\beq}{\begin{equation}}
\newcommand{\eeq}{\end{equation}}
\newcommand{\bea}{\begin{eqnarray}}
\newcommand{\eea}{\end{eqnarray}}

\begin{document}

\title{Larmor precession and Debye relaxation of single-domain magnetic nanoparticles}

\author{Zs. J\'{a}nosfalvi$^1$\footnote{corresponding author: janosfalvi.zsuzsa@atomki.mta.hu}, J. Hakl$^1$ and P.F. de Ch\^{a}tel$^{3,2,}$}
\affiliation{Institute of Nuclear Research, P.O.Box 51, H-4001 Debrecen, Hungary
\\$^2$Department of Physics, New Mexico State University, Las Cruces, NM 88003, USA
\\$^3$Institute of Metal Research, CAS, Shenyang 110016, P.R. China}

\begin{abstract} 
The numerous phenomenological equations used in the study of the behaviour of single-domain magnetic nanoparticles are described and some issues clarified by means of qualitative comparison. To enable a quantitative \textit{application} of the model based on the Debye (exponential) relaxation and the torque driving the Larmor precession, we present analytical solutions for the steady states in presence of circularly and linearly polarized AC magnetic fields. Using the exact analytical solutions, we can confirm the insight that underlies Rosensweig's introduction of the "chord" susceptibility for an approximate calculation of the losses. As an important consequence, it can also explain experiments, where power dissipation for both fields were found to be identical in "root mean square" sense. We also find that this approximation provides satisfactory numerical accuracy only up to magnetic fields for which the argument of the Langevin function reaches the value $2.8$.\end{abstract}


\maketitle 

\section{Introduction}
\label{sec_intro}
Enduring interest in colloidal dispersions of ferro- and ferrimagnetic nanoparticles is fueled by their  applications. Colloids of small particles of iron oxide are used in magnetic resonance imaging (MRI) \cite{na} as contrast agents and in hyperthermia treatment as heat-generating media \cite{hergt}. Of the numerous applications of the exchange of energy between the magnetic particles and the fluid around them we are interested herein hyperthermia. The importance of this process lies in the transport of the energy the magnetic particles absorb from an external magnetic field into the cancer tissue. Dissolving the particles in a colloid is useful in targeting them to the tumour, where they may get anchored in inferior arteries. We assume that further movement of the particles, other than rotation, and the flow of the surrounding blood inside the tumour are not important. Even with this constraint, the process of energy exchange is complex and has been described in terms of numerous phenomenological models, treated in a variety of approximations. In an earlier paper \cite{chatel}, we have presented the analytic solutions of the Landau-Lifshitz-Gilbert (LLG) equation for isotropic single-domain particles in AC and circularly polarized field; later, Nándori and Rácz \cite{nandori} included uniaxial anisotropy as well.

The LLG equation describes the movement of a magnetic dipole in the presence of a time-dependent magnetic field. Measurements on ferrofluids provide data on the magnetization, i.e. the average over a large number of single-domain nanoparticles. Solutions of the LLG equation for various initial states are not sufficient to calculate the magnetization; the question of averaging is still left open. In this paper we report the solution of an equation of motion of the magnetization, which contains explicitly the torque driving the dipoles to the direction of the field. In itself, this torque leads to an exponential approach to the direction of a static external field; the Debye relaxation. Debye has studied the movement of electric dipoles carried by molecules \cite{debye}. In the case of magnetic dipoles, the coupling of the angular momentum to the magnetic momentum brings inevitably the Larmor torque into the equation of motion. This term is familiar from the Bloch equation in the literature of magnetic resonance \cite{bloch}. In a comparison of eight phenomenological equations of motion for magnetic moments of ferro- or ferrimagnetic nanoparticles dispersed in a nonmagnetic medium Berger \textit{et al.} \cite{berger} listed the equation combining the Larmor torque and the Debye relaxation under the name "modified Bloch equation". This is the equation we have solved, analyzed and compared with the LLG equation in this work.

In the next section, we give an overview of the various equations of movement used for the description of the simultaneous effect of external torques and relaxation, showing the place of the modified Bloch equation among them. En route, we give the shortest derivation of the Landau-Lifshitz-Gilbert \cite{landau, gilbert} equation from the Landau-Lifshitz equation. In Section \ref{sec_circ} we give an analytical solution of the modified Bloch equation for circularly polarized magnetic field. Surprisingly, the case of a linearly polarized field is more challenging; the analytical solution we can give in Section \ref{sec_lin} is not valid for arbitrarily strong fields. 

\section{Equations of motion and simplifications}
\label{sec_eqs}
The behaviour of single-domain ferro- or ferrimagnetic  nanoparticles in an external magnetic field $\bm{B}$ has much in common with that of atomic or nuclear magnetic moments. The torque on a magnetic moment $\bm{\mu}$,

\begin{equation}
	\bm{T} = \bm{\mu} \times \bm{B}
	\label{torque}
\end{equation}

\noindent determines the equation of motion of the angular moment, $d\bm{L}/dt=\bm{T}$. The gyromagnetic relation, $\bm{\mu} =\gamma \bm{L}$ enables a closed equation for $\bm{\mu}$ which, applied to the magnetic moment of unit volume, provides the equation of motion of the magnetization,

\begin{equation}
	d\bm{M}/dt = \gamma\bm{M} \times \bm{B}.
	\label{dynamicsM}
\end{equation}

\noindent Here $\gamma$ is the gyromagnetic ratio. The magnetization of materials we have in mind in this work is due to the spin of electrons, accordingly $\gamma= - 1.76 \times 10^{11}$ Am$^2$/Js. 

The vector product in eq.(\ref{dynamicsM}) implies that any change of the magnetization is perpendicular to $\bm{M}$, that is, the modulus of $\bm{M}$ remains constant. Also, if $\bm{B}$ is constant, $d( \bm{M}\cdot\bm{B})/dt=0$, that is, the angle between the two vectors is constant. The only motion satisfying these conditions is a precession of $\bm{M}$ around $\bm{B}$. The angular velocity of the magnetization in this \textit{Larmor precession} is $\bm{\omega_L} = \gamma\bm{M}\times\bm{B}/M_{\bot}$, where $M_{\bot}$ is the projection of $\bm{M}$ on the plane perpendicular to $\bm{B}$. The Larmor frequency is defined as a positive quantity, $\omega_L=|\gamma| B$. To reduce the potential energy, $U=-\bm{M}\cdot\bm{B}$, the contribution to $d\bm{M}/dt$ due to relaxation must have a component parallel with $\bm{B}$. Landau and Lifshitz \cite{landau} have chosen a "damping term", which evidently achieves this, being proportional to $\bm{M}\times(\bm{M}\times\bm{B})=(\bm{M}\cdot\bm{B})\bm{M}-M^2\bm{B}$. The \textit{Landau-Lifshitz equation} of motion is then

\begin{equation}
\frac{d\bm{M}}{dt}=\gamma[\bm{M}\times\bm{B}+\alpha M^{-1}\bm{M}\times(\bm{M}\times\bm{B})].
	\label{landau-lifshitz}
\end{equation}

\noindent The coefficient $\alpha$ goes under the name of "the dimensionless damping coefficient", which is something of a misnomer, because addition of the "damping term" evidently \textit{enhances} the motion of $\bm{M}$:

\begin{equation}
\left|\frac{d\bm{M}}{dt}\right|=\gamma|\bm{M}\times\bm{B}|(1+\alpha^2)^{1/2}.
	\label{enhances}
\end{equation}

\noindent Gilbert's approach \cite{gilbert} is closer to the notion of friction, as it subtracts from the Larmor torque a torque proportional to $-d\bm{M}/dt$. This is reminiscent of friction in linear motion, where a force opposite to the velocity $d\bm{r}/dt$ is introduced into the equation of motion. The analogy allows a derivation of the \textit{Gilbert equation},

\begin{equation}
\frac{d\bm{M}}{dt}=\gamma[\bm{M}\times\bm{B}-\eta\mu_0\bm{M}\times d\bm{M}/dt],
	\label{gilbert}
\end{equation}

\noindent by adding a Rayleigh dissipation function to the Lagrangian which describes the Larmor precession of the magnetic moment.

A comparison of the Landau-Lifshitz and Gilbert equations reveals that $d\bm{M}/dt$ (i) is perpendicular to $\bm{M}$ in both equations and (ii) in the former it consists of two mutually perpendicular terms while in the latter this is not the case. Observation (ii) implies that decomposition of the damping term in the Gilbert equation into components parallel and perpendicular to that of the Landau-Lifshitz equation will offer a direct comparison of the two. In fact, the Gilbert equation itself provides a decomposition into components which delivers the desired transformation. Multiplying both sides of eq.(\ref{gilbert}) by $\bm{M}$ and taking (i) in account yield

\begin{equation}
\bm{M}\times\frac{d\bm{M}}{dt}=\gamma[\bm{M}\times(\bm{M}\times\bm{B})+\eta\mu_0 M^2 d\bm{M}/dt].
	\label{modgilbert}
\end{equation}

\noindent Substituting this result in the last term of the Gilbert equation and rearranging terms lead to the \textit{Landau-Lifshitz-Gilbert} (LLG) \textit{equation},

\begin{equation}
\frac{d\bm{M}}{dt}=\gamma(1+\alpha^2)^{-1}[\bm{M}\times\bm{B}-\alpha M^{-1}\bm{M}\times (\bm{M}\times\bm{B})],
	\label{llg}
\end{equation}

\noindent where $\alpha=\gamma\eta\mu_0 M$. Clearly, for $\alpha \ll 1$ the Landau-Lifshitz equation is a good approximation, but for the general case the $(1+\alpha^2)^{-1}$ factor is essential to eliminate the non-physical implications of the Landau-Lifshitz equation pointed out by Kikuchi \cite{kikuchi} and Gilbert \cite{gilbert}. Also, due to this factor, Gilbert's damping term \textit{reduces} the motion of $\bm{M}$:

\begin{equation}
\left|\frac{d\bm{M}}{dt}\right|=\gamma\left|\bm{M}\times\bm{B}\right|(1+\alpha^2)^{-1/2}.
	\label{reduces}
\end{equation}

\noindent Strictly speaking, the effect of the Landau-Lifshitz and Gilbert damping coefficients cannot be described with a relaxation time, because the relaxation they stand for is not exponential. If $\bm{B}$ is a constant field, pointing in the $z$ direction, the solution of the LLG equation is $M_z = M \tanh(\alpha\tilde{\omega}_L t)$, with $\tilde{\omega}_L = \omega_L/(1+\alpha^2)^{-1/2}$. 

Shliomis \cite{shliomis1974} has suggested that under well-defined conditions the equation of exponential relaxation,

\begin{equation}
\frac{d\bm{M}}{dt}=-\frac{\bm{M}-\bm{M}_{eq}}{\tau},
	\label{debye}
\end{equation}

\noindent should suffice to describe the behaviour of a colloid of magnetic nanoparticles. Here, $\bm{M}$ is the average magnetization of the particles, 

\begin{equation}
\bm{M}_{eq}=M_S {\cal L}\left( \frac{\mu_0 HM_d V}{kT} \right)\hat{\bm{e}}_H,
	\label{meq}
\end{equation}

\noindent where $V$ is the particle volume, $\hat{\bm{e}}_H=\bm{H}/H$ is the unite vector pointing along $\bm{H} = \bm{B}/\mu_0$ and $M_S$ is the saturation magnetization of the colloid $M_S=\phi M_d$, $\phi$ by being the volume fraction and $M_d$ is the magnetization in the single-domain magnetic nanoparticle. In the Debye relaxation equation, especially when applied in magnetic resonance, as well in the LL and G equations, it is customary to use $\mu_0 \bm{H}$, rather than $\bm{B}$. Gilbert pointed out in the Appendix of ref.\cite{gilbert} that $\bm{H}$ is not limited to the externally applied field and he gives the definitions of the demagnetizing field, the exchange fields, the anisotropy fields and the magnetoelastic fields in terms of the concomitant energies.

The Langevin function, ${\cal L}(x)=\coth(x)-1/x$, gives the magnitude of the magnetization in thermal equilibrium. Note that $M_{eq}$ must be an ensemble average and consequently so is $\bm{M}$. In this respect, in the context of superparamagnetic resonance or hyperthermia, eq.(\ref{debye}) is more expedient than the LLG equation. In the latter case, having found the possible solutions of the equation of motion, one has to face the issue of the appropriate  weighted average of the associated energy losses. 

Clearly, the parameter $\tau$ in eq.(\ref{debye}) is a proper relaxation time. In a stationary field, where $\bm{M}_{eq}$ is also constant, the solution of this equation for the component of $\bm{M}$ along $\bm{H}$ is

\begin{equation}
(M(t)-M_{eq})=(M(0)-M_{eq})\exp(-t/\tau).
	\label{soldebye}
\end{equation}

\noindent In reference~\onlinecite{shliomis2003}, Shliomis names eq.(\ref{debye}) the \textit{Debye relaxation equation}. 

Recently, Cantillon-Murphy \textit{et al}. \cite{cantillon-murphy} have published a thorough analysis of the implications of Debye relaxation equation, eq.(\ref{debye}). Apart from the approximation inherent in the exclusion of the gyromagnetic torque they also applied Rosensweig's chord susceptibility \cite{rosensweig2002},

\begin{equation}
\chi_{ch}=\frac{M_S}{H_0} {\cal L}\left( \frac{\mu_0 H_0 M_d V}{kT} \right)
	\label{chord}
\end{equation}

\noindent instead of the Langevin function appearing in eq.(\ref{meq}).

\begin{figure}
  	\resizebox{80mm}{!}{
			\includegraphics{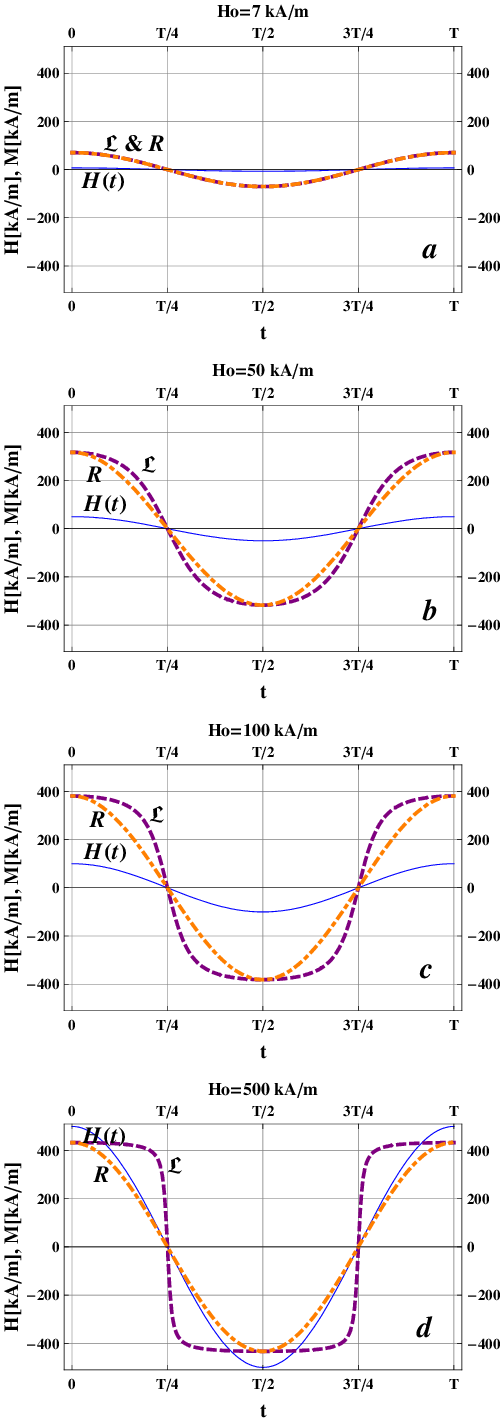}}
		\caption{\label{fig:chord}(colour online) The characteristic change of the equilibrium magnetization within a periodic driving cycle as given by the Langevin function ($\cal{L}$, eq. (\ref{meq})) and Rosensweig's chord susceptibility ($R$, eq. (\ref{chord})) for different $H_0$ amplitude magnetizing field $H(t) = H_0\cos(2\pi t/T)$. At low $H_0$ field amplitudes the two curves are identical (\textbf{\textit{1a}}). At increasing $H_0$ field amplitudes only the extrema remain identical.}
\end{figure}

\begin{figure}
  	\resizebox{80mm}{!}{
			\includegraphics{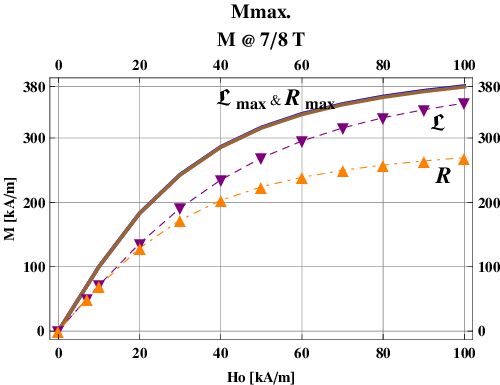}}
		\caption{\label{fig:amplis} (colour online) Equilibrium magnetization $({\cal L})$ and approximate equilibrium magnetization $(R)$ calculated for $t = 7T/8$ and $t = T$ $(\cal{L}_{\max} \& R_{\max})$ plotted against the amplitude $H_0$ of the AC magnetic field. At $t = T$ the two are identical by definition of $\chi_{ch}$, but at $t = 7T/8$ the relative difference between the ${\cal L}$ and $R$ curves increases with $H_0$, reaching $20\%$ at $100$ kA/m.}
\end{figure}

Equation (\ref{debye}) implies that in a field $H(t) = H_0 \cos(2 \pi t/T)$ the relaxation pulls the magnetization towards the equilibrium magnetization, which is oscillating in time. Figure \ref{fig:chord} shows this equilibrium magnetization determined by the Langevin function $({\cal L})$ and the Rosensweig's chord susceptibility $(R)$. By definition of the latter, the two curves meet at $H = 0$ and $H = H_0$. In the limit of strong external field $(\mu_0 H_0 M_d V/kT \gg 1)$, Fig.\ref{fig:chord}(d), the ${\cal L}$ curve is flipping between the extrema, while the $R$ curve has sinusoidal characteristic. The magnetization does not exceed the saturation magnetization $M_S$, even for field $H_0>M_S$.  In the weak field limit, Fig.\ref{fig:chord}(a), where hyperthermia is applied, there is no difference. Figures \ref{fig:chord}(b) and \ref{fig:chord}(c) show the transition between the two behaviours. It is clear that when the field reaches a value where $M_{eq}/M_S > 0.8$ (Figs. \ref{fig:chord}(c) and \ref{fig:chord}(d)), the magnetization calculated with the chord susceptibility substantially deviates from the correct function. 

The dependence of the discrepancy between the two curves on the amplitude of the AC field is shown in Fig.\ref{fig:amplis}. In the low-field limit there is no discrepancy, because $\chi_{ch}=\chi$ and $M_{eq}=\chi H$, but it is visible at $20$ kA/m and increases with increasing field. Beyond $H_0 = 100$ kA/m the relative difference remains constant at about $20\%$, which is too large to use the approximation in calculations of the magnetization. In Session \ref{sec_lin} we show that this does not disqualify the chord susceptibility in calculation of the energy loss.

In sections \ref{sec_circ} and \ref{sec_lin} we present analytical solutions to the Debye relaxation equation enriched with the Larmor torque,

\begin{equation}
\frac{d\bm{M}}{dt}=\mu_0\gamma \bm{M}\times\bm{H} -\frac{\bm{M}-\bm{M}_{eq}}{\tau},
	\label{debyewlarmor}
\end{equation}

\noindent with $\bm{M}_{eq}$ as given in eq.(\ref{meq}), $\bm{M}$ is the average of magnetization of the single-domain magnetic particles and $\bm{H}$ is the external magnetic field. In the calculations the effect of internal fields (crystalline anisotropy and demagnetization fields) is not considered. We will use these results to assess the effects of the torque on the steady-state solutions under linear and circular polarization of the AC magnetic field. As we take into account the curvature  of the Langevin function, we can also discuss the implications of the chord susceptibility.

\section{Circular polarization}
\label{sec_circ}
In this section we give the analytical solution of eq.(\ref{debyewlarmor}) for a rotating magnetic field. The first term, representing the Larmor torque, spoils the separation of Cartesian components, which has enabled the derivation of the solution, eq.(\ref{soldebye}), found in Section \ref{sec_eqs} for the Debye relaxation equation. For a rotating magnetic field, $H_x=H_0 \cos(\omega t); H_y=H_0 \sin(\omega t); H_z=0$, the coupled equations to be solved are as follows:

\begin{eqnarray}
	\frac{dM_x}{dt}&=&-\mu_0\gamma M_z H_0 \sin(\omega t)-\frac{1}{\tau}\left(M_x-M_{eq}(H_0)\cos(\omega t)\right);\nonumber\\
		\frac{dM_y}{dt}&=&\mu_0\gamma M_z H_0 \cos(\omega t)-\frac{1}{\tau}\left(M_y-M_{eq}(H_0)\sin(\omega t)\right);
		\nonumber\\
			\frac{dM_z}{dt}&=&\mu_0\gamma (M_x H_0 \sin(\omega t)-M_y H_0 \cos(\omega t))-\frac{M_z}{\tau}.\label{ccde}
\end{eqnarray}

\begin{figure}
  	\resizebox{80mm}{!}{
			\includegraphics{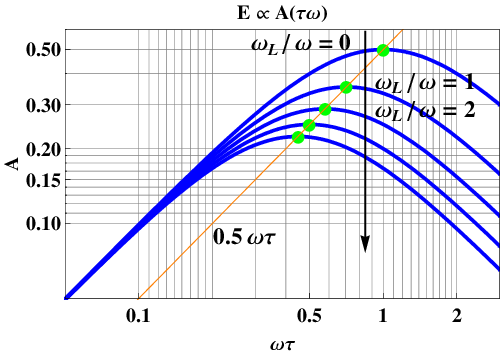}}
		\caption{\label{fig:A} (colour online) The dependence of $A\propto E$, the energy loss, on $\omega\tau$ for various $\omega_L/\omega$ ratios from $0$ to $4$. See eq. (\ref{energyCirc}).}
\end{figure}

\noindent To handle the entanglement of the three components of $\bm{M}$, it will prove to be convenient to apply the series of transformations that enabled us in a previous work \cite{chatel} to solve this set of equations for free precession, i.e. in the absence of relaxation $(1/\tau = 0)$. Three transformations create a coordinate system, which rotates as dictated by free Larmor precession in the rotating field. First, a rotation around the $z$ axis by an angle $\omega\tau$ makes the $xy$ plane follow the magnetic field, then the $z$ axis is turned by an angle $\Theta$ into the direction of the total angular velocity $\bm{\Omega} =\bm{\omega}+\bm{\omega}_L$, and finally a rotation around this new, $z'$ axis by an angle $\Omega\tau$ drives the $x'y'$ plane to rotate together with the magnetization vector. The
description is reminiscent of an Euler transformation and indeed the product of the three matrices representing the rotations listed above, is of the form of the canonical Euler transformation \cite{varshalovich} with the replacements $\alpha\leftrightarrow\Omega t, \beta\leftrightarrow -\Theta\  \text{and}\  \gamma\leftrightarrow -\omega t$. As the axis of the Larmor precession is the magnetic field, which rotates in the plane perpendicular to the $z$ axis, $\bm{\omega}$ is perpendicular  to $\bm{\omega_L}$ and $\Omega=\sqrt{\omega^2+\omega^2_L}$. Also, it follows from this configuration that $\sin\Theta=\omega_L/\Omega$ and $\cos\Theta=\omega/\Omega$. In what follows, the transformation matrix will be used in the form

\begin{widetext}
\begin{equation}
\underline{\underline{\bm{O}}}=\frac{1}{\Omega}
\begin{pmatrix}
	\omega\cos\omega t\cos\Omega t + \Omega\sin\omega t\sin\Omega t & \omega\sin\omega t\cos\Omega t - \Omega\cos\omega t\sin\Omega t & -\omega_L\cos\Omega t\\
		\omega\cos\omega t\sin\Omega t - \Omega\sin\omega t\cos\Omega t & \omega\sin\omega t\sin\Omega t + \Omega\cos\omega t\cos\Omega t & -\omega_L\sin\Omega t\\
	\omega_L\cos\omega t & \omega_L\sin\omega t & \omega
	\end{pmatrix}.
	\label{trmx}
\end{equation}
\end{widetext}

\noindent To find the derivative of the transformed magnetization vector $\bm{M}'$, we need the derivative of the matrix $\underline{\underline{\bm{O}}}$:

\begin{equation}
\frac{d\bm{M}'}{dt}=\frac{d\underline{\underline{\bm{O}}}\bm{M}}{dt}=\frac{d\underline{\underline{\bm{O}}}}{dt}\bm{M}+\underline{\underline{\bm{O}}}\frac{d\bm{M}}{dt}.
\label{differentiation}
\end{equation}

\noindent Substituting here $\frac{d\bm{M}}{dt}$ from eq.(\ref{debyewlarmor}), we find that the contribution of the Larmor torque to $\underline{\underline{\bm{O}}}\frac{d\bm{M}}{dt}$ cancels $\frac{d\underline{\underline{\bm{O}}}}{dt}\bm{M}$, as it should, leaving the following differential equations for the transformed magnetization:

\begin{equation}
\tau\frac{d\bm{M}'}{dt}=-\underline{\underline{\bm{O}}}\bm{M}+\underline{\underline{\bm{O}}}
\begin{pmatrix}
M_{eq}(H_0)\cos(\omega t)\\
M_{eq}(H_0)\sin(\omega t)\\
0
\end{pmatrix}.
	\label{transmag}
\end{equation}

\noindent The first term on the right-hand side gives trivially $-\bm{M}'$, the second one can be found applying eq.(\ref{trmx}), to find the differential equations for the three components of the transformed magnetization vector. Ultimately, the analytical solution for the transformed magnetization vector is 

\begin{widetext}
\begin{eqnarray}
	M'_x(t)&=&\left[ M'_x(0)-M_{eq}(H_0)\frac{\omega}{\Omega}\frac{\cos\delta}{\sqrt{1+(\Omega\tau)^2}} \right]\exp(-t/\tau)+M_{eq}(H_0)\frac{\omega}{\Omega}\frac{\cos(\Omega t-\delta)}{\sqrt{1+(\Omega\tau)^2}};\nonumber\\
	M'_y(t)&=&\left[ M'_y(0)+M_{eq}(H_0)\frac{\omega}{\Omega}\frac{\sin\delta}{\sqrt{1+(\Omega\tau)^2}} \right]\exp(-t/\tau)+M_{eq}(H_0)\frac{\omega}{\Omega}\frac{\sin(\Omega t-\delta)}{\sqrt{1+(\Omega\tau)^2}};\nonumber\\
	M'_z(t)&=&\left[ M'_z(0)-M_{eq}(H_0)\frac{\omega_L}{\Omega} \right]\exp(-t/\tau).
	\label{analtrans}
\end{eqnarray}
\end{widetext}

Here $\delta$ is defined by $\sin\delta=\Omega\tau/\sqrt{1+(\Omega\tau)^2},\ \cos\delta=1/\sqrt{1+(\Omega\tau)^2}$. Since the exponentially decaying terms are of no interest for applications on time scales exceeding $\tau$ by several orders of magnitude $(t/\tau \gg 1)$ and we seek  the steady-state solution in the laboratory frame, we shall drop the exponential terms. The inverse transformation is easily carried out with the transpose of matrix (\ref{trmx}). A compact form of the final result is

\begin{equation}
\bm{M}(t)=M_{eq}(H_0)\frac{\omega}{\Omega^2}\frac{1}{1+(\Omega\tau)^2}
\begin{pmatrix}
\omega\cos(\omega t)+\Omega^2\tau\sin(\omega t)\\ \omega\sin(\omega t)-\Omega^2\tau\cos(\omega t)\\ -\omega_L
\end{pmatrix}.
	\label{analfinal}
\end{equation}

\noindent Note that $|\bm{M}(t)|=M_{eq}(H_0)\frac{\omega}{\Omega} \frac{1}{\sqrt{1+(\Omega\tau)^2}},$ indicating that the interplay of Larmor precession and relaxation pushes the magnitude of the magnetization, which is time-independent, below the thermal equilibrium value in a \textit{static} field $H_0$. The energy loss per cycle is easily calculated,

\begin{eqnarray}
E&=&-\mu_0 \int_0^{0+2\pi/\omega} \bm{M}\cdot\frac{d\bm{H}}{dt}\ dt\nonumber\\
&=&2\pi\mu_0 M_{eq}(H_0)H_0 \frac{\omega\tau}{1+(\Omega\tau)^2}.
	\label{energyCirc}
\end{eqnarray}

\noindent This Debye-type dependence on $\Omega\tau$ is familiar in low-field magnetic resonance \cite{garstens1955} and was shown by Garstens \cite{garstens1954} to be exact in the limit of vanishing static field. 

The basic functional dependence of $E$ and $H_0, \omega$ and $\tau$ can be complicated if $H_0$ is too large to use a susceptibility to calculate $M_{eq}(H_0)$. Separating $M_{eq}$, we can analyze the term $H_0\omega\tau/(1 + \Omega^2\tau^2)$ as a function of $\omega\tau$ and $\omega_L/\omega$. As $\tau$ is an elusive quantity, not very easy to control experimentally, we have plotted $A=\omega\tau/(1+\Omega^2\tau^2)$ as a function of $\omega\tau$ under the condition that $\omega_L/\omega = constant$ for various values of $\omega_L/\omega$, see Fig. \ref{fig:A}. The maxima of these functions are seen to occur where $(dA/d(\omega\tau))_{\omega_L/\omega}=0$. Indeed this conditional derivative vanishes at $(\omega\tau)_{max}=[1+(\omega_L/\omega)^2]^{-1/2}$, where $A((\omega\tau)_{max})=\omega\tau/2$, which is reflected in the line connecting the maxima in Fig. \ref{fig:A}.

\begin{figure}
  	\resizebox{80mm}{!}{
			\includegraphics{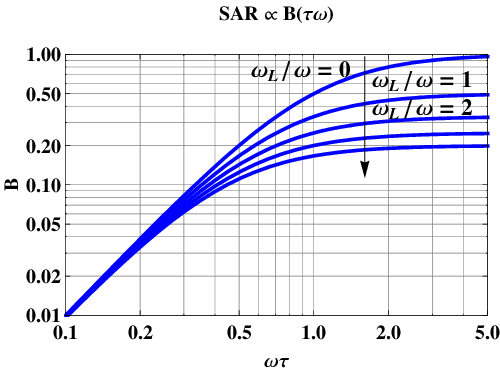}}
		\caption{\label{fig:B}(colour online) The dependence of $B \propto SAR$, the specific absorption rate on $\omega\tau$ for various $\omega_L/\omega$ ratios from $0$ to $4$. See eq. (\ref{B}).}
\end{figure}

The basic functional dependence of $E$ on $H_0$, $\omega$ and $\tau$ is dictated by the term $\frac{H_0 \omega\tau}{1+\Omega^2\tau^2}$. Noting, that $\omega_L^2\propto H_0^2$, it has maximum either as a function of $H_0$, $\omega$, $\tau$ or products $\omega\tau$ and $\omega_L\tau$. To facilitate comparison with Garstens and Kaplan' \cite{garstens1955} result for the frequency dependence of $E$, we select the dimensionless parameters $\omega\tau$ and $\omega_L / \omega$ and define $n=1+\omega_L / \omega$ and $A=\frac{\omega}{\Omega}\frac{\Omega\tau}{1+(\Omega\tau)^2}$. Figure \ref{fig:A} shows $A$ (which is proportional to $E$) as a function of $\omega\tau$ at various value of $n$. The energy loss is seen to increase with $\omega\tau$ up to a maximum at $1/n$, the value of $A$ at the maxima being $1/(2n)$. It is remarkable how the interplay of relaxation and Larmor precession suppresses the energy loss, in the case of circular polarization. 

The specific absorption rate (SAR) defined as energy loss per second and per kg of the colloid, is the gauge of energy losses relevant to applications:

\begin{equation}
SAR \doteq \frac{E\omega}{2\pi\rho}=\frac{\mu_0 M_{eq}(H_0)H_0}{\rho}\frac{\omega^2\tau}{1+(\Omega\tau)^2}.
\label{SAR}
\end{equation}

This is the figure of merit for colloids to be used in hyperthermia. In this application the week magnetic fields are quite weak ($\sim 10^5$ A/m) and the temperature not too low ($> 300$K), the ferrofluid is far from saturation and the usage of a susceptibility is justified, so that $M_{eq}(H_0)\propto H_0$. Equation (\ref{SAR}) can be visualized then by the function 

\begin{equation}
B=\frac{(\omega_L\tau)^2\omega\tau}{1+(\omega\tau)^2+(\omega_L\tau)^2}\ ,
\label{B}
\end{equation}

\noindent which is plotted for various values of $\omega_L\tau$ in Fig.\ref{fig:B}. In this case, the maxima can be found without any condition at $(\omega\tau)_{max}=[1+(\omega_L\tau)^2]^{1/2}$ and the value of $B$ at the maxima is given by $B((\omega\tau)_{max})=\frac{(\omega_L\tau)^2-1}{2\omega\tau}$, which is the equation of the curve connecting the maxima in Fig. \ref{fig:B}.

\section{Linear polarization}
\label{sec_lin}
To find the equation of motion for the magnetization subjected to a magnetic field oscillating along the $z$ axis,  the following vectors have to be inserted into eq.(\ref{debyewlarmor}): $\bm{H}(t)=(0,0,H_0\cos(\omega t)), \bm{M}\times\bm{H}=\frac{\omega_L}{\mu_0|\gamma|}(M_y\cos(\omega t),-M_x\cos(\omega t),0)$ and $\bm{M}_{eq}(\bm{H})=\left(0,0,\phi M_d {\cal L} \left(\frac{\mu_0 M_d V H_0}{kT}\cos(\omega t)\right)\right)$, where $\omega_L=\mu_0|\gamma|H_0$. The matrix of the transformation that will eliminate the Larmor term of eq.(\ref{debyewlarmor}) depends on the sign of the gyromagnetic ratio. For both cases we can write the matrix as 

\begin{equation}
\underline{\underline{\bm{O}}}_{\pm}=
\begin{pmatrix}
\pm\sin g & \cos g & 0\\
-\cos g & \pm\sin g & 0\\
0 & 0 & 1
\end{pmatrix},
	\label{lintrmx}
\end{equation}

\noindent where the upper sign equals that of $\gamma$ and $g(t)=(\omega_L/\omega)\sin(\omega t)$. The equation of motion for $\bm{M}'(t)=\underline{\underline{\bm{O}}}_{\pm}\bm{M}(t)$ is

\begin{widetext}
\begin{equation}
\frac{d\bm{M}'(t)}{dt}=\frac{d\underline{\underline{\bm{O}}}_{\pm}}{dt}\bm{M}+\underline{\underline{\bm{O}}}_{\pm}\frac{d\bm{M}}{dt}=\frac{d\underline{\underline{\bm{O}}}_{\pm}}{dt}\bm{M}\pm\mu_0|\gamma|\underline{\underline{\bm{O}}}_{\pm}(\bm{M}\times\bm{H})-\frac{1}{\tau}\underline{\underline{\bm{O}}}_{\pm}\left( \bm{M}-\bm{M}_{eq}(\bm{H}) \right).
	\label{magnew}
\end{equation}
\end{widetext}

\noindent Taking into account that $dg/dt=\omega_L\cos(\omega t)$ (irrespective of the sign of $\omega_L$), the first two terms in eq.(\ref{magnew}) cancel each other. The equation of motion for $\bm{M}'(t)$ in terms of the new coordinate system is easy to find, as $\underline{\underline{\bm{O}}}_{\pm}\bm{M}=\bm{M}'$, by definition, and $\bm{M}_{eq}(\bm{H})$ has only a $\mathit{z}$ component, which is evidently not affected by $\underline{\underline{\bm{O}}}_{\pm}; M_{z'} = M_z$. What remains in the transformed frame is the set of equations of motion

\begin{eqnarray}
\frac{dM_x'}{dt}&=&-\frac{1}{\tau}M_x';\nonumber\\
\frac{dM_y'}{dt}&=&-\frac{1}{\tau}M_y';\nonumber\\
\frac{dM_z'}{dt}&=&-\frac{1}{\tau}\left[M_z'-M_S {\cal L}\left( \zeta \cos(\omega t) \right)\right].
	\label{linsol}
\end{eqnarray}

\noindent The solution for the first two equations must be the familiar exponential relaxation to zero,

\begin{equation}
M_{x,y}'(t)=M_{x,y}'(0)\exp(-t/\tau).
	\label{translinsolxy}
\end{equation}

\noindent Since $\underline{\underline{\bm{O}}}_{\pm}$ separates the $xy$ plane and the $z$ axis, the transformation back to the laboratory frame can be done in two dimensions. As $g$ vanishes at $t=0$, we have $M_x(0)=-M'_y(0)$ and $M_y(0)=M'_x(0)$, whence

\begin{eqnarray}
\bm{M}(t)&=&
\begin{pmatrix}
\pm M_y(0)\sin g+M_x(0)\cos g\\
M_y(0)\cos g \mp M_x(0)\sin g 
\end{pmatrix}
\exp(-t/\tau)\nonumber\\
&=&
M_{\bot}(0)
\begin{pmatrix}
\cos[g\mp\varphi(0)]\\
\mp\sin[g\mp\varphi(0)]
\end{pmatrix}
\exp(-t/\tau),
	\label{linsolxy}
\end{eqnarray}

\noindent where $\varphi(0)$ is the azimuth angle of $\bm{M}(0)$ and $M_{\bot}(0)=\sqrt{M^2_x(0)+M^2_y(0)}$. We can determine the time dependence of the angular velocity:

\begin{eqnarray}
\varphi&=&\tan^{-1}\frac{M_y}{M_x}=\tan^{-1}\left( \mp\tan\left( \frac{\omega_L}{\omega}\sin(\omega t)\mp\varphi(0) \right) \right)\nonumber\\
&=&\mp\frac{\omega_L}{\omega}\sin(\omega t)\mp\varphi(0);\nonumber\\
\frac{d\varphi}{dt}&=&\mp\omega_L\cos(\omega t),
	\label{angle}
\end{eqnarray}

\noindent with the upper sign for $\gamma>0$ and the lower sign for $\gamma<0$. Equation (\ref{angle}) describes a Larmor precession whose frequency is following the time dependence of the magnetic field (note that $\omega_L$ is defined as the Larmor frequency at $H=H_0$, so that $\omega_L\cos(\omega t)$ gives the Larmor frequency at $H=H(t)$). Whether the time dependence of the precession frequency is observable will depend on the magnitude of the projection of $\bm{M}$ on the $xy$ plane, which, according to eq.(\ref{linsolxy}), decays exponentially,

\begin{equation}
\sqrt{M^2_x(t)+M^2_y(t)}=M_{\bot}(0)\exp(-t/\tau).
	\label{linlenght}
\end{equation}

\noindent If the relaxation time is very short (like in hyperthermia, where $\omega\tau\approx10^{-4}$), the magnetization will be fully aligned along the $z$ axis before the magnetic field undergoes a substantial change, let alone a change of sign. Of course, to observe the time dependence given in eq.(\ref{linsolxy}), we also need a Larmor frequency larger than the frequency of $H$, that is, $\omega_L > \omega$. These conditions are met in hyperthermia, at $H_0 = 200$ kA/m the Larmor frequency is about $40$ GHz, whereas $\omega$ is of the order of $100$ kHz.

 The last equation in (\ref{linsol}), which is also the equation of motion of $M_z$, can be reduced to a single (though not simple) integral. Multiplying both sides with $e^{t/\tau}$ and reordering terms lead to

\begin{equation}
\frac{d(e^{t/\tau}M_z)}{dt}=\frac{M_S}{\tau}e^{t/\tau}{\cal L}(\zeta\cos(\omega t)).
	\label{identity_a}
\end{equation}

\noindent To avoid the difficult integral, for $\zeta\ll 1$ one can take the susceptibility instead of the Langevin function, i.e., the first term in the Taylor series of ${\cal L}$, i.e. the Curie susceptibility, instead of the function ${\cal L}$. In fact, the integrals can be done for higher order terms also (Ref.~\onlinecite{gradshteyn}, p.228). As the Taylor series of $\coth x$ necessary here (Ref.~\onlinecite{gradshteyn}, p.42) is only valid for $|x|<\pi$, the same restriction, $\cos \zeta < \pi$, is valid on the series of the Langevin function,

\begin{equation}
{\cal L}(\zeta\cos(\omega t))=\sum_{m=1}^{\infty}\frac{2^{2m}}{(2m)!}B_{2m}(\zeta\cos(\omega t))^{2m-1},
	\label{taylorLangevin}
\end{equation}

\noindent where the $B_{2m}$ are Bernoulli numbers as defined and listed in Ref.~\onlinecite{gradshteyn},  p.1040 and p.1045, respectively. Substituting the series (\ref{taylorLangevin}) into eq.(\ref{identity_a}),

\begin{equation}
e^{t/\tau}M_z=\frac{M_S}{\tau}\sum_{m=1}^{\infty}\frac{2^{2m}}{(2m)!}B_{2m}\zeta^{2m-1}\int e^{t/\tau}\cos^{2m-1}\omega t\ dt.
	\label{identity_d}
\end{equation}

\noindent Now it is straightforward to find a solution to the equation of motion for $M_z$ which is analytical, albeit in the form of an infinite series :

\begin{widetext}
\begin{equation}
M_z=\frac{2M_S}{\tau}\sum_{m=1}^{\infty}\frac{1}{m}B_{2m}\zeta^{2m-1}\sum_{k=0}^{m-1}\frac{(1/\tau)\cos((2k+1)\omega t)+(2k+1)\omega\sin((2k+1)\omega t)}{(m-1-k)!(m+k)!\left[(1/\tau)^2+(2k+1)\omega^2\right]}.
	\label{infiniteseries}
\end{equation}
\end{widetext}

\begin{figure}
  	\resizebox{80mm}{!}{
			\includegraphics{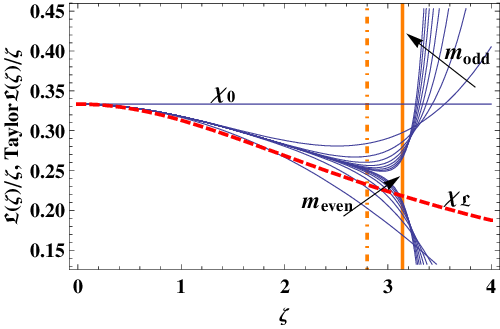}}
		\caption{\label{fig:convergence}(colour online) Convergence of the Taylor series expansion of ${\cal L} (\zeta)/\zeta$ representing the sum term in eq.(\ref{effsolchord}) up to odd and even labels increasing in the direction of the arrows up to $m_{max}=17$. The solid line belongs to $m_{max}=1 (\chi_0)$. The thick dashed line shows $\chi_{{\cal L}}={\cal L}(\zeta)/\zeta$.}
\end{figure}

\noindent There are three features of this solution that can be revealed without evaluating the series: (i) the exponential factors have disappeared, meaning that the function describes a steady state, (ii) keeping only the $m = 1$ term in the series the well-known result for the frequency-dependent Curie susceptibility $M_z=\chi_0 H_0 [\cos(\omega t)+\omega\tau \sin(\omega t)]/[1+(\omega\tau)^2]$ with

\begin{equation}
\chi_0=\frac{\mu_0 M_d^2 V\phi}{3kT}
	\label{curiesolz}
\end{equation}

\noindent emerges and (iii) the double summation is essentially a Fourier  series, of which only the $k = 0$ sine terms are needed for the calculation of losses, because $M_z$, multiplied by $-dH(t)/dt=H_0 \omega\sin(\omega t)$, will be integrated over a cycle. 

Keeping only the $k = 0$ terms in the second sum in eq.(\ref{infiniteseries}) amounts to eliminating the higher harmonics in the time dependence of the magnetization, which are generated due to the substantial nonlinearity of the Langevin function beyond $x\approx 0.5$. It follows then that for any value of $H_0$ a susceptibility can be found, which will provide the exact energy loss on the assumption that the magnetization is

\begin{equation}
M_z^{eff}=2M_S\sum_{m=1}^{\infty}\frac{1}{(m!)^2}B_{2m}\zeta^{2m-1}\frac{\cos(\omega t)+\omega\tau\sin(\omega t)}{1+(\omega\tau)^2}.
	\label{effsolz}
\end{equation}

\noindent Looking back at Fig.\ref{fig:chord}(d), it becomes now clear that abrupt changes of the magnetization triggered by the flipping of the equilibrium magnetization will not influence the calculated energy loss, because they are represented by high-frequency terms in $M_z$.

The effective magnetization of eq.(\ref{effsolz}) may be looked upon as a fictitious magnetization, which, if realized, would give the correct losses without reproducing the correct hysteresis  loop. Instead, the shape of the hysteresis loop and its frequency dependence are exactly the same as those one gets at low fields, where the static magnetization is proportional to the magnetic field. Equation (\ref{effsolz}) confirms then the insight that underlies Rosensweig's introduction of the chord susceptibility [Ref. \cite{rosensweig2002}, eq.(\ref{dynamicsM})] and enables the calculation of the "field dependence" of a fictitious susceptibility. In fact, the susceptibility relevant to the energy loss, implicitly defined by $M_z^{eff}=\chi_{loss}H_0[\cos(\omega t)+\omega\tau \sin(\omega t)]/[1+(\omega\tau)^2]$ with

\begin{equation}
\chi_{loss}=2\frac{\mu_0 H_0 M_d^2 V \phi}{kT}\sum_{m=1}^{\infty}\frac{1}{(m!)^2}B_{2m}\zeta^{2(m-1)},
	\label{effsolchord}
\end{equation}

\noindent is \textit{not field} dependent (that would contradict the definition of the susceptibility as a limes at $H\rightarrow0$), but \textit{amplitude} dependent, through $\zeta$, which is proportional to $H_0$. Note that $B_2=1/6$, so that $\chi_{loss}=\chi_0$, the Curie	susceptibility, for $\zeta\ll 1$. 

In practice, the summation in eq.(\ref{effsolchord}) will be limited to a finite index $m_{max}$. The rapid increase of the Bernoulli numbers with $m$ is a concern, but Fig.\ref{fig:convergence} shows that taking $m_{max}=17$, the convergence is reliable up to $\zeta=2.8$. The converged values of ${\cal L}(\zeta)/\zeta$ represent the amplitude dependence of the fictitious susceptibility $\chi_{loss}$. Note that while the deviation of $\chi_{chord}$ from $\chi_0$ is increasing substantially, the difference between the exact $\chi_{loss}$ and $\chi_{chord}$ is small and remains constant within about $5\%$ even for $2 < \zeta < 2.8$. Hence for $\zeta > 1$ calculating the energy loss using $\chi_0$ is not reliable, but close enough results can be obtained using the $\chi_{chord}$ susceptibility. 

We have calculated the energy loss per cycle, using the formula

\begin{widetext}
\begin{equation}
E=-\mu_0 \int^{t_0+2\pi/\omega}_{t_0} \bm{M}\cdot\frac{d\bm{H}}{dt}dt=-\mu_0 \int^{t_0+2\pi/\omega}_{t_0} M_z^{eff}\cdot\frac{dH_z}{dt}dt=\pi\mu_0 H_0 \chi_{loss}(\zeta)\frac{\tau\omega}{1+(\tau\omega)^2}.
	\label{energyLin}
\end{equation}
\end{widetext}

\begin{figure}
		\resizebox{80mm}{!}{
			\includegraphics{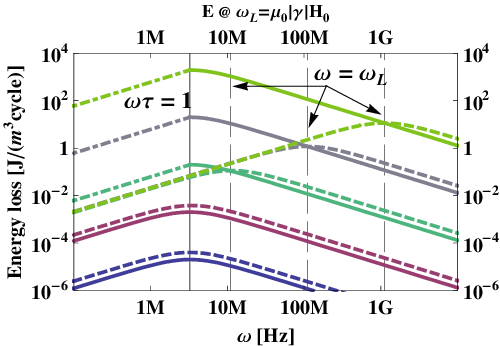}}
		\caption{\label{fig:energy}(colour online) Energy loss as a function of frequency $\omega$, parameterized by field strength $H_0=\{10^4, 10^3, 10^2, 10^1, 10^0\}$ A/m from top to bottom. Solid lines correspond to linear polarization, dashed lines to circular. Vertical lines are at resonance $\omega\tau$ and at the Larmor frequency belonging to $H_0=\{10^4, 10^3, 10^2\}$ A/m from right to left. The calculation were done at magnetite nanoparticle of radius $5$ nm and $\tau = 3\times 10^{-7}$ s.}
\end{figure}

\noindent The results are given in Figs.\ref{fig:energy} and \ref{fig:SAR} and will be discussed in section \ref{sec_summary} below. Comparing the $\omega\tau$ dependence of $E$ in both polarized cases, prompt can be realized that frequency dependence of energy dissipation in circularly and linearly polarized cases can be equivalent by a factor of $2$, if $n=1+\omega_L/\omega=1$ and the linear approximation of $M_{eq}$ is taken for both polarized cases. This can also be the qualitative interpretation of experimental results of O.O. Ahsen \textit{et al.}\cite{ahsen}, where identity of both polarized fields, having the same "root mean square" magnitude, were stated for power dissipation.

At high driving field amplitudes $(\zeta\gg 1)$ the energy absorption per cycle can be estimated as follows. For $\omega \ll 1/\tau$ eq.(\ref{linsol}) implies that $M_z$ lags behind the target $M_S {\cal L} (\zeta\cos(\omega t))$ value by time $\tau$, which yields for the energy absorption $E \approx 4\mu_0 M_S H_0 \omega\tau$.  For  $\omega \gg 1/\tau$ according to eq.(\ref{linsol}) the driving magnetic field causes only saw-tooth curve like oscillation of magnetization around the zero value with amplitude $\approx M_S \pi/2\omega\tau$. The energy absorption per cycle in this case is $E\approx 4\mu_0 M_S H_0/\omega\tau$.

\begin{figure}
  	\resizebox{80mm}{!}{
			\includegraphics{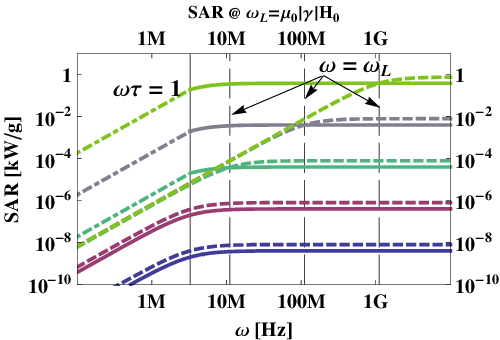}}
		\caption{\label{fig:SAR}(colour online)  Specific absorption rate as a function of frequency $\omega$. Notation as under FIG. \ref{fig:energy}.}
\end{figure}
\section{Summary and conclusions}
\label{sec_summary}
 
In section \ref{sec_eqs} we have attempted to clarify the relationship between a number of models, defined by the appropriate equations of motion and, in some cases, the approximations used to deal with the equations. Concerning the latter, we have analysed the suggestion of Rosensweig \cite{rosensweig2002} to introduce a susceptibility (the chord susceptibility) dependent on the amplitude of the AC magnetic field, instead of following the nonlinear Langevin function. The large differences between the exact and approximate time dependence of the magnetization shown in Fig. \ref{fig:chord} suggest that using the chord susceptibility would lead to serious underestimation of the SAR. 

The analytic solutions of the modified Bloch-Bloembergen equation for circular and linear polarization, presented in sections \ref{sec_circ} and \ref{sec_lin}, respectively, are in excellent agreement with the numerical solutions. In the case of linear polarization our solution is in the form of a Fourier series. Inserting this into the integral that gives the energy loss, it becomes clear that the features of the magnetization that in Section \ref{sec_eqs} seemed to disqualify the chord susceptibility in fact do not contribute to the SAR. This insight enables an estimation of the range where the chord susceptibility can be used. This range contains the parameters used in hyperthermia.

The frequency dependence of the energy dissipation is shown in Fig. \ref{fig:A}. The maxima are seen to decrease as the ratio $\omega_L/\omega$ increases and to shift to lower frequency when the relaxation time is kept constant. The specific absorption rate (SAR, Fig. \ref{fig:B}) shows saturation with increasing $\omega\tau$ and it decreases with increasing $\omega_L/\omega$ . The equivalence of linear and circularly polarized fields can be verified (taking account of the factor of two due to the two linear components amounting to the circular one) for $\omega_L/\omega << 1$  and $\mu_0 H_0 M_d V/kT<<1$.

To illustrate the difference between circular and linear susceptibilities mentioned in Section \ref{sec_lin}, in Fig. \ref{fig:energy} we show the energy loss per cycle $(E)$ as function of the driving angular frequency  $(\omega)$ in five different values of the magnetic field amplitude $(H_0)$ for both polarizations. The curves are log-log plots of eq. (\ref{energyCirc}) and (\ref{energyLin}) for the circular (dashed lines) and linear (full lines) polarization, respectively. It is assumed that in eq. (\ref{energyCirc}) $M_{eq}$ and in eq. (\ref{energyLin}) $\chi_{loss}$ is proportional with $H_0$. The range of magnetic fields covers four orders of magnitude.

The dependence of $E$ on $\omega$ is quite simple in both cases. It is easily seen that if $\omega_L << \omega$, for very low frequencies $E\propto\omega$ and for very large frequencies $E\propto\omega^{-1}$. Hence the tent-like shapes in the log-log plot of Fig.\ref{fig:energy}, which can also understood as follows. For $\omega \ll 1/\tau$ eq.(\ref{linsol}) implies that $M_z$ lags behind the target $M_S {\cal L} (\zeta\cos(\omega t))$ value by time $\tau$, which yields for the energy absorption $E \approx 4\mu_0 M_S H_0 \omega\tau$.  For  $\omega \gg 1/\tau$ according to eq.(\ref{linsol}) the driving magnetic field causes only saw-tooth curve like oscillation of magnetization around the zero value with amplitude $\approx M_S \pi/2\omega\tau$. The energy absorption per cycle in this case is $E\approx 4\mu_0 M_S H_0/\omega\tau$.

For linear polarization, $dE/d\omega\propto 1 - \omega^2\tau^2$, the maxima in between the two straight lines do not depend on the magnetic field, i.e. the Larmor frequency. However, for linear polarization $dE/d\omega\propto 1 + \omega_L^2\tau^2 - \omega^2\tau^2$, which shifts the maxima towards higher frequencies. Indeed, the plots do not show any difference in the positions of the maxima in the curves for weak magnetic fields, while in strong fields the maxima for circular polarization shift to higher frequencies in steps amounting to about one order of magnitude. This is easily explained taking in account that $\omega_L\propto H_0$. Likewise, the almost identical straight lines running up to the points marked with $\omega=\omega_L$ in Figs. \ref{fig:energy} and \ref{fig:SAR} indicate that in that range $\omega<<\omega_L\propto H_0$. Note that in these calculations we have chosen $\gamma=8.82\times 10^{10}$ Am$^2$/Js.

As to the relative merits of linear and circular polarization, we come to the conclusion that followed from our work in the Landau-Lifshitz-Gilbert frame \cite{chatel}: at the frequencies relevant to hyperthermia the linear option will produce more energy loss. To come to this conclusion from Figs. \ref{fig:energy} and \ref{fig:SAR}, it is important to realize that the constant splitting between the thick and broken lines seen in the lower, right-side part of the pictures represents the factor of $2$ mentioned above in connection with Figs. \ref{fig:A} and \ref{fig:B}. We see the circular choice loosing even this apparent advantage as we cross the diagonal along which $\omega=\omega_L$. In the upper left corner, at low frequencies and magnetic fields stronger than $10$ kA/m, the circular polarization is useless for hyperthermia. It is gratifying to see that this conclusion confirms the finding of Berger et al. \cite{berger} that various phenomenological approach can serve equally well in the description of different aspects of magnetism.

\section*{Acknowledgement}

We wish to express special thanks to I. N\'andori and to colleagues in the lab for their valuable critical advices and continuous support. The authors acknowledge support from the Hungarian Scientific Research Fund (OTKA) No.101329. This work was supported by the TAMOP 4.2.2.A-11/1/KONV-2012-0036 project, which is co-financed by the European Union and European Social Fund.

\end{document}